\documentclass{article}

\usepackage{bm}
\usepackage{amssymb}
 \usepackage{amsthm}
 \usepackage{amsmath}
 \usepackage{cite}
 \usepackage{float}

\usepackage{arxiv}
\usepackage[utf8]{inputenc} 
\usepackage[T1]{fontenc}    
\usepackage{hyperref}       
\usepackage{url}            
\usepackage{booktabs}       
\usepackage{amsfonts}       
\usepackage{nicefrac}       
\usepackage{microtype}      

\usepackage{graphicx}

\title{Asymptotic Incidence Rate estimation of SARS-COVID-19  via a Polya process scheme: a comparative analysis in Italy and European Countries}

\author{ \href{https://orcid.org/0000-0000-0000-0000}{\includegraphics[scale=0.06]{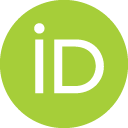}\hspace{1mm}Filippo Carone~Fabiani}
\\
	Department of Engineering and Applied Science\\
	University of Bergamo\\
	Bergamo, 24127 \\
	\texttt{filippo.caronefabiani@unibg.it} \\
}

\renewcommand{\headeright}{}
\renewcommand{\undertitle}{}

\hypersetup{
pdftitle={Incidence rate estimation of SARS-COVID-19  via a Polya process scheme: a comparative analysis in Italy and European},
pdfsubject={q-bio.NC, q-bio.QM},
pdfauthor={Filippo Carone~Fabiani},
pdfkeywords={COVID-19, Polya urn, Negative Binomial, Incidence Rate}
}

\begin{document}
\maketitle

\begin{abstract}
	During an ongoing epidemic, especially in the case of a new agent, data are partial and sparse,  also affected by external factors as for example climatic effects or  preparedness and response capability of healthcare structures.
	Despite that we showed how, under some universality assumptions, it is possible to extract strategic insights by modelling the pandemic trough a probabilistic Polya urn scheme.
In the Polya framework,  we provide both the distribution of infected cases and the asymptotic estimation of the incidence rate,  showing that data are consistent with a general underlying process    at different scales. 
Using European confirmed cases and diagnostic test data on COVID-19 , we provided an extensive comparison among European countries and between Europe and Italy at regional scale, for both the two big waves of infection.
We globally estimated an incidence rate in accordance with previous studies. On the other hand, this quantity could play a crucial role as a proxy variable for an unbiased estimation of the real incidence rate, including symptomatic and asymptomatic cases. 
\end{abstract}


\section{Introduction}

On December 2019, a novel coronavirus (SARS-CoV-2)-infected pneumonia (COVID-19) was first identified  in Wuhan, Hubei (China). Due to the extensive spreading of the infection, on March 11, the World Health Organization (WHO) declared it a pandemic \cite{WHO}. Since then, more than 80 millions of confirmed cases and about 2 millions of deaths have been reported worldwide. Just in Europe, confirmed and deaths have been respectively about 25 and 0.5 millions of cases, mostly concentrated in Russia, France, Italy, U.K. and Spain.  
Since the outbreak of the pandemic many national and international health organizations  have collected daily data about the COVID-19 pandemic, although following different policy-making in terms of specific informations provided, temporal and geographical aggregation of data and  efficiency of tests. For example, Spain, Germany, Netherlands and Sweden have provided only weekly data and few of them have provided aggregated data on a small regional scale. There are also countries that  started later to provide tests or, in some cases, they didn't report any data about performed tests (Albania, Moldova, Montenegro). Moreover, due to the emergency caused by the ongoing epidemic,  helplessness in detecting concurrent causes of death   and, in some cases, delays or corruptions in data reporting, have affected the database on regional and national level. Last but not least, due to the excess load of the healthcare systems and to the unknown nature of the pathogen, direct counting of the total number  of infected patients (confirmed, asymptomatic and pauci-symptomatic cases) is impeded \cite{Pape, Cow}. This leads to a biased estimation of key indexes, as the case fatality rate (CFR) and the infection fatality rate (IFR) \cite{Ma,Nish,Woo}, usually used during a pandemic to measure the lethality of the incipient infection.
Due to the above limitations, performing a rigorous analysis  to assess the patterns of the epidemic, is difficult, mostly in the case of a new pathogen. 
 Nevertheless, despite the scarcity 
and uncertainty of the available data, a simple analysis of empirical curves of European confirmed cases  suggests some universality in the epidemic spreading. Multiple waves of infection and closed values of key indexes \cite{Rus,Ver,Fer} observed in countries under a wide social and geographical conditions, suggest the presence of distinctive features of the infection  according to  same underlying process at different scales. 

Here, we present an extensive  analysis,  of  the COVID-19 pandemic spread, at national and regional scales,  comparing results from 37 European countries and 21 Italian regions, for both the two big observed waves of infection.
We  showed that the dynamics of the COVID-19 susceptible-infection system can be appropriately modelled by a generalized Polya urn scheme, previously used in epidemiology to model disease transmission for infectious diseases like SARS or smallpox \cite{lloy} and to implement more general transmission models \cite{HAY}. The probabilistic Polya  urn scheme \cite{Polya1, Polya2} describes the stochastic process of drawing with reinforcement from an urn,  a number of  balls with two different colours (here labelling  healthy and infected cases) and it is able to explain  important features of the COVID-19 pandemic's scenario.  So far as is known, in the limit of long time, it directly provides, both the probability density function (PDF) of   infected cases and an  estimation of the asymptotic composition $\rho_{\infty}$ of the urn. 
 
In order to implement  the Polya scheme, we applied a simplified multiple waves approach in which each wave of infection is considered an independent process. For each selected wave, confirmed cases observations were used  to  fit the Negative Binomial PDFs that govern the Polya drawings, for each selected country and region. We statistically tested PDFs resulting from our fitting, obtaining   about $74\%$ of  successful PDFs for European countries and about $64\%$ for Italian regions, over both the waves of infection.  Within the same wave, estimated parameters resulted very close and consistent with  some general underlying behaviour for a wide range of external conditions.
As in  mean field approximation spirit, for each wave of the infection, and for each group of data (European and Italian) we assumed the same initial conditions and the same constant reproduction number. As a consequence,  both national and regional data can be  considered as independent series of trials related to the same process at different scales. This enabled us  to estimate  the asymptotic composition $ \rho_{\infty}$ of the urn, as the mean  of the Beta PDF, which governs the  limiting distribution of the sample average of a Polya process. In order to estimate the Beta PDF we used the time series of   the empirical  ratio between  the cumulative number of confirmed cases and the cumulative number of performed  diagnostic tests.
The entire procedure has been repeated for both the two big infection waves of the pandemic, for each  group of the European countries and for each group of the Italian regions.
The comparative results between European and Italian spread are consistent with the assumption of universality stated for our procedure, enabling us  to provide an asymptotic estimation of the incidence rate (IR), by adopting $\rho_{\infty}$  as a global variable. The computed $\rho_{\infty}$ of confirmed cases represents a very important feature in the pandemic dynamics, in that it could play a crucial role as a proxy variable to estimate  the IR of the total infected cases (symptomatic and asymptomatic) on the total susceptible population, hence the total number of cases, which in turns represents a key value to get strategic information required for the public health policy-making processes \cite{Lip}.

\section{Methodology}
\label{sec:headings}
 
Hereinafter, we described how  the spreading of the COVID-19 infection can be  ascribed to a contagion process based on the Polya urn scheme.
In its basic version, a single urn is considered, initially containing a number $N$ of balls: $w$ white balls plus $b$ black balls. A ball is drawn at random and then replaced together with a number $d$ of balls of the same color. The parameter $d$ simulates the reproducibility number, it means the average number  of people which are infected when in touch with a sick person. The procedure is repeated $n$ times, with $n$ unlimited.
 In such a scheme it is known that, for $n \rightarrow \infty$ and large $N$, the probability mass function of drawing $m$ white balls after the $n$-th draw  can be approximated by a Negative Binomial PDF $N\!B(r,p)$, with parameters $r=\frac{w}{d}$,  and $p=\frac{N}{N+nd}$ \cite {Tee}.   Furthermore, once denoting the  process  by  the indicator function $I_n$ (equal to $1$ or $0$ if   the drawn ball is respectively white or not), the fraction of the total number of white balls inside the urn after $n$-th draw can be written as:  
\begin{equation}
\rho_n=\frac{\rho_0+n \delta Z_n}{1+n \delta}
\end{equation} 
 where  $\delta=\frac{d}{N}$ being the normalized $d$, $\rho_0=\frac{w}{N}$ is the  initial fraction of white balls,  and
 \begin{equation}
 Z_n=\frac{1}{n}\sum\limits_{t=1}^{n}I_t
 \end{equation}
is the sample average of the process.
  It can be proved that, setting  $\theta=\frac{N}{d}-r$,  $\rho_n$ and $ Z_n$  are martingale, both  converging almost surely as $n \rightarrow \infty$  to a random variable distributed according to a Beta $B\big(r,\theta\big)$ with a  mean value $\rho_{\infty}$  coinciding  with the initial fraction  $\rho_0$ of white balls \cite{Fel,KOT1,KOT2}.  Note that the above parametrization explicitly links the distribution of the random process with the distribution of its sample average.

The above properties can be extended to more general schemes: balls labelled with an arbitrary number of colors, different replacement rules ($d$) \cite{KOT2}, or different number $m$ of drawn balls at once (multiple drawings)\cite{Aou} and \cite{May-Ru}. In the latter case it can be also proved that the total fraction of white balls  $ \rho_n$ are still converging  to a random variables  with a Beta-like distribution. 
This supports the idea to describe confirmed cases data by a Polya process in whatever time or geographical format they have been aggregated.
In order to assure suitable conditions for implementing the entire scheme, we need to set the assumptions and statements  below. Hereafter we will use the apex $g=E,I$ to label quantities respectively related to the group of the European countries or the Italian regions  and the apex $\nu=1,2$  to label quantities related to the first or the second waves of infection. 

\subsection{Assumptions and statement}

\begin{enumerate}
\item We assume an ideal efficiency of the diagnostic  tests, it means  tests are $100\%$ sensitive  for detecting COVID-19 infections.

\item  Regular and persistent features, common to many countries  with different levels of disease severity, highlight a general underlying dynamic of the spread of the disease.  As for many countries, although with different social and geographical conditions, the presence of different peaks observed in the curve of confirmed cases (Fig.13)  suggests to model the disease spread by multiple waves of infection, considering each of them  as a single independent infection process. We concentrated our analysis on the two biggest waves observed in most of the considered countries.

\item We applied this multiple waves scheme    describing each of  the two waves by a single uni-modal distribution. In order to separate dataset of each country we selected  time intervals around the two highest peaks, whose endpoints coincide  with the lowest values before and after a peak. For practical calculation, we assume these values represent respectively the onset and the close of a single wave of infection although in many cases the there is an overlap between consecutive waves.

\item We restricted our analysis excluding abrupt change in dynamical conditions, and applying our procedures to consecutive waves of infection. In other words, we assume  the pandemic, after a finite time interval,  has exhausted its actual  pathogenic load, due to natural causes or to containment rules adopted. This allows us to deal  with  asymptotic quantities in the limit of large time $t$.

\item We denote by  $c_{n,i}^{g,\nu}$ and $s_{n,i}^{g,\nu}$ observations related respectively to the $n$-th daily number of confirmed case and $n$-th daily number of performed diagnostic test for the $i$-th European country ($g=E$ and  $i=1,...,37$ ) or for the $i$-th Italian region ($g=I$ and  $i=1,...,21$) during the first ($\nu=1$) or second ($\nu=2$) wave of infection.
Identifying  the white balls with the variable $c_{n,i}^{g,\nu}$ and  the number of drawn balls, at each step $n$, with the daily tests  $s_{n,i}^{g,\nu}$, we describe the infection spread by a Polya process, assuming that each time series $c_{n,i}^{g,\nu}$   follows a Negative Binomial  distribution $N\!B(r_{i}^{g,\nu},p_{i}^{g,\nu})$, with parameters $r_{i}^{g,\nu}$ and $p_{i}^{g,\nu}$ defined as above for each $i$, $g$ and $\nu$.


\item We  applied  our procedure  separately to   the set of  national data  and to the regional one so as to analyse and compare the  spreading of COVID-19 at different scales  in terms of  number of population and  geographical size. We repeated the procedure for the first and the second wave of infection.
\item Based on the assumption 2, we assumed that, for each wave of infection, the group of European countries was approximately in the  same local conditions: actually they have approximately the same initial ratio  between infected and susceptible cases ($ \rho_{0,i}^{E,\nu}\simeq \rho_{0}^{E,\nu}$), and the same normalized  reproduction number ($ \delta_i^{E,\nu} \simeq \delta^{E,\nu}$), independently from the $i$-th considered country. This implies that, for large $n$, since we can rewrite $ r_{i}^{E,\nu}\equiv \frac{\rho_{0,i}^{E,\nu}}{\delta_i^{E,\nu}}$, and $p_{i}^{g,\nu} \equiv \frac{1}{1+n\delta_i^{E,\nu}}$, also $ r_{i}^{E,\nu} \simeq r^{E,\nu} \equiv \frac{\rho_{0}^{E,\nu}}{\delta^{E,\nu}}$ and $p_i^{E,\nu} \simeq p^{E,\nu} \equiv \frac{1}{1+n\delta^{E,\nu}}$.
In order to account these conditions, we assumed that, at fixed $\nu$, the Negative Binomials of assumption 5 can be seen as  Negative Binomial compound distributions  $N\!B(r_i^{E,\nu},p_i^{E,\nu})$ in which the characteristic  parameters are both normally distributed with a small variance, such that $ r_{i}^{E,\nu}$  and $ p_{i}^{E,\nu}$ can be approximated by their mean values: $ r_{i}^{E,\nu}\simeq r^{E,\nu}$ and  $ p_i^{E,\nu} \simeq p^{E,\nu}$.
Finally, for each wave of infection,   European countries can be represented by  independent urns, each of them containing a different total number of balls, corresponding to the different population, but providing  the same 
Negative Binomial compound distribution  $N\!B(r^{E,\nu},p^{E,\nu})$. 

 \item Let define $R_{n,i}^{E,\nu} \equiv \frac{\sum\limits_{t=1}^{n}c_{t,i}^{g,\nu}}{\sum\limits_{t=1}^{n}s_{t,i}^{g,\nu}}$ the ratio between cumulative cases and cumulative diagnostic tests. In our scheme each $R_{n,i}^{E,\nu}$ replaces  the sample average $Z_{n,i}^{E,\nu}$ of eq.2 for each process. Inspired by general results \cite{Aou,May-Ru}, we assume that each $R_{n,i}^{E,\nu}$ follows a Beta limiting distribution $B\big(r_i^{E,\nu},\theta_i^{E,\nu}\big)$.  On the other hand, according to the  assumption 7 they can reduce to
  the same $B\big(r^{E,\nu},\theta^{E,\nu}\big)$, since $ \theta_{i}^{E,\nu}  \simeq \theta^{E,\nu}\equiv \frac{1}{\delta^{E,\nu}}-r^{E,\nu}$.

\item The previous assumptions 7-8 were stated also at regional scale,  considering data related to the group of the  Italian regions with different population but with  the same $\rho_0^{I,\nu}$ and the same $\delta^{I,\nu}$, for all $i$. 
   
\end{enumerate}

Assumption 7 is the hard hypothesis of our procedure and it will verified a-posteriori by succeeding results.

\subsection{Infection rate estimation based on universality assumptions }

For our calculations confirmed cases and  diagnostic tests data of  the 37 accessible  European countries were collected from 24 February 2020 to 30 January 2021. Data have reported by Humanitarian Data Exchange (HDX) database \cite{HDX}, that provides national data worldwide.
  For the Italian case, we consider all  the  21   regions using data  provided  by the Italian Civil Protection Agency (CPA) database \cite{PCI} on daily time scales.  
Under the above assumptions  we fitted the $N\!B(r_{i}^{g,\nu},p_{i}^{g,\nu})$ PDFs  for each of the 37 European selected countries and for each of the 21 Italian regions.
For properly selecting observations for each wave of infection and for each country, we proceeded  as described in assumption 3. In most of the countries,  a multiple waves path is recognizable from daily confirmed cases  curve. Although shifted by a time delay,  all confirmed cases  curves show a first isolated wave followed by a second  or a third  big wave overlapping each other and a number of smaller peaks that seem to be not ascribable to random fluctuations. 
We consider only the two biggest waves,  locating  the onset and close points  simply considering the minima  around each peak.   
The maximum likelihood estimation algorithm was used to compute the fitting parameters $r_{i}^{g,\nu}$ and $p_{i}^{g,\nu}$ . For practical calculation, we adopted the \texttt{fitdist} function  implemented in Matlab, training the models with the confirmed cases data $c_{n,i}^{g,\nu}$. 
In order to select PDFs that successfully fit confirmed cases data, we  performed both Kolmogorov-Smirnov  (KS-test)  and Chi-square test ($\chi^2$-test)  using respectively the \texttt{kstest} and the \texttt{chi2gof} Matlab functions. As a result, over both the two waves, we were able to select a subset of European countries (about $74\%$ of successful PDFs)  and a subset of Italian regions(about $64\%$ of successful PDFs) that passed both the tests.
 To assess the validity of our procedure we must to justify the assumption 7.  
To this aim, we first computed sample means and variances of the estimated  parameters $r_i^{g,\nu}$ ($\mu_r^{g,\nu}$ and $\sigma_r^{g,\nu}$ ) and $p_i^{g,\nu}$  ( $\mu_p^{g,\nu}$ and $\sigma_p^{g,\nu}$), then,  under normality assumption we statistically tested  whether each variance $\sigma_*^{g,\nu}$ was  significantly different from zero.  
 A  t-test was  performed for each $\sigma_*^{g,\nu}$, using \texttt{ttest} Matlab function. Successful results, from  all of the  performed tests, allow us to approximate  $r_i^{g,\nu}$ and $p_i^{g,\nu}$, according to assumption 7 and accounting for universality stated in assumption 2 and 7.
Once the distributions of confirmed cases have been proved, the problem of estimating the asymptotic value  $\rho_{\infty}$ reduces to  compute the mean of the asymptotic distributions  $B\big(r^{g,\nu},\theta^{g,\nu}\big )$. In fact, according to assumption 8, for each waves, we can consider data from each European country, or  Italian region,  as $i$ different  sequences of trials of a process with different population but characterized by  the same Beta PDF. 
This implies that, for each group  $g$ and for each wave of infection $\nu$, we can assume  $R_{n,i}^{g,\nu}$ are governed by the same $B\big(r^{g,\nu},\theta^{g,\nu}\big)$. As a result we were able to fit  the four  Beta PDFs  by using the set of the $i$ limiting values $R_{\infty,i}^{g,\nu}= \lim_{n\to\infty}  R_{n,i}^{g,\nu}$   (see Fig.14-15). A cut-off value has been also introduced to establish a finite size convergence criterion. However, in most of the cases $R_{\infty,i}^{g,\nu}$ were obtained directly by the ratio computed on  the end points of each waves of infection, since they can be considered far enough  to represent the limit for large times,  where the stability of  $R_{n,i}^{g,\nu}$ is reached. 
Also in this case,  maximum likelihood estimation algorithm and  \texttt{fitdist}  function was adopted to estimate the parameters $ r^{g,\nu}$ and $ \theta^{g,\nu}$, for each  Beta PDFs. KS-test and $\chi^2$-test were also successfully conducted   to assess the performance of the estimated Beta. 
In order to assure the consistency of the whole procedure we tested whether each  $ r^{g,\nu}$, estimated by $B\big(r^{g,\nu},\theta^{g,\nu}\big)$, was  significantly different from respective $\mu_r^{g,\nu}$ estimated by $N\!B(r_{i}^{g,\nu},p_{i}^{g,\nu})$. For each group $g$ and each wave of infection  $\nu$ we implement a one-sample two tailed t-test,  under the null hypothesis $ \mu_r^{g,\nu}=r^{g,\nu} $.   A two-sample two tailed t-test to simultaneously compare couples of $\mu_r^{g,\nu}$ with crossed $g$ and $\nu$  suggesting the invariance of $ r^{g,\nu}$ under change of $\nu$ or $g$ and similarity behaviour at different scales.
Finally  an estimation of the mean value $\rho_{\infty}^{g,\nu}$ was obtained. All results are provided in the next sessions.

\section{Results}
\subsection{European countries}
 For our calculation we restricted to the countries  with total population grater than one million, selecting 40 of the 48 countries from European continent, reported in the  HDX database (see Tab.1-2). We also discarded countries that did not provide diagnostic test data at all (Albania, Montenegro, Moldova). We preprocessed remaining 37 European countries,  first  removing  inconsistent data (outliers and negative data) than selecting only observations corresponding to performed diagnostic tests.
Wherever possible, data were collected on daily time scale, otherwise we smoothed weekly  data, $c_{n,i}^{g,\nu}$ and $s_{n,i}^{g,\nu}$, as in the case of Germany, Spain, Netherlands, France and Ukraine. Due to different starting day of the pandemic,  delays or corruptions in data reporting, four countries (Bulgaria, Cyprus, Czechia, North Macedonia ) did not provided sufficient data for the first wave of infection.  As a result, we obtain two sets composed by  33 countries for the first wave and  by 37 for the second wave.
 Fitting procedures were applied on the above two groups and both KS-test and $\chi^2$-test with a confidence level $\alpha=0.01$, were performed to select successful PDFs, whose p-values were reported in Tab.1-2.   Resulting $i=27$ successful  $N\!B(r_i^{E,1},p_i^{E,1})$ PDFs for the first wave are showed in Fig.1-3, where we reported fitted PDFs  (Fig.1) and their relative CDFs (Fig.2-3).  The same procedures  were performed for the second wave (Fig.4-6) obtaining $i=28$ successful  $N\!B(r_i^{E,2},p_i^{E,2})$.  Fitting parameters $r_i^{E,\nu}$ and $p_i^{E,\nu}$ of all successful PDFs  were reported in Tab.1-2 (first wave) and Tab.4-5 (second wave).   In Fig.2-3 and Fig.5-6,  is also reported  the graphical comparison between fitted  CDFs and empirical distribution functions, computed with a $95\%$ confidence interval ($\%$ c.i.). 
By using the successful subset of data  we  computed sample means and variances ($\mu_r^{E,\nu}$, $\sigma_r^{E,\nu}$  and $\mu_p^{E,\nu}$, $\sigma_p^{E,\nu}$).
Excluding the two outliers values coming from Ukraine and Poland, for the first wave, we obtained $\mu_r^{E,1}=1.48$ ($95\%$ c.i.: $1.34 - 1.62 $)  and  $\mu_p^{E,1}=3.17 \cdot 10^{-2}$  ($95\%$ c.i.: $2.06 \cdot 10^{-2} - 4.28 \cdot 10^{-2} $) (see Tab.3).
Similar results were obtained for the second wave: $\mu_r^{E,2}=1.68$ ($95\%$ c.i.: $1.54-1.82 $)  and  $\mu_p^{E,2}=0.04 \cdot 10^{-2}$  ($95\%$ c.i.: $0.03 \cdot 10^{-2} - 0.05 \cdot 10^{-2} $). 
 For each wave of infection we assessed the sharpness of the normal distributions $r_i^{E,\nu}$ and $p_i^{E,\nu}0$ by using two   one-sample two-tailed t-tests,  with a significance level $\alpha=0.05$, under the null hypothesis that respectively $\sigma_r^{E,\nu}=0$ and $\sigma_p^{E,\nu}=0$. All tests were  successful, with a p-values within the interval (0.08 - 0.6).
 Focusing on parameters $r_i^{E,\nu}$ we also note that $\mu_r^{E,1}$ is very closed to $\mu_r^{E,2}$ within the c.i., suggesting it is invariant also under change of the wave $\nu$ considered. In order to verify  the latter hypothesis ($\mu_r^{E,1}=\mu_r^{E,2}$)  a two-sample two-tailed t-test, with  a significance level $\alpha=0.05$, was also performed with success, by using \texttt{ttest2} function, and obtaining a p-value 0.09.
 On the contrary, concerning, the sample means $\mu_p^{E,\nu}$ they can't directly compared with a target value since the parameters $p^{E,\nu}$ can't be directly estimated from $B\big(r^{g,\nu},\theta^{g,\nu}\big)$. On the other hand, $\mu_p^{E,\nu}$  strongly decreases  passing from the first wave to the  second one. The deviation observed   can be due by different values of  $d$ trough $\delta^{E,\nu}$: increasing values of $\delta^{E,\nu}$ imply decreasing value of $p_i^{E,\nu}$ and of the relative sample means $\mu_p^{E,\nu}$.  
All these results support our assumptions (2 and 7-9)  although a more detailed analysis is prevented since $\delta^{g,\nu}$ cannot be directly estimated from $p_i^{g,\nu}$ or $\theta^{g,\nu}$ . 
Once the convergence of $R_{n,i}^{E,\nu}$ was reached, for each selected  country, asymptotic values of the ratio $R_{\infty,i}^{E,\nu}$ were used to fit the two Beta PDFs $B\big(r^{E,\nu},\theta^{E,\nu}\big)$, (see Fig.7). KS-test  and $\chi^2$-test were successfully performed on both the distributions. Fitting parameters $r^{E,\nu}$ and $\theta^{E,\nu}$ related to the Beta distributions were estimated and reported in Tab.3.  We found  a value of $1.80$ ($95\%$ c.i.: $1.07 - 3.02$)  for $r^{E,1}$  and  $1.95$ ($95\%$ c.i.: $ 1.10 - 3.46$) for $r^{E,2}$, showing that, within the c.i., $r^{E,\nu}$ are consistent with the sample means obtained from estimated Negative Binomials.  In order to complete the comparison between parameters satisfying the initial condition in assumption 7-9, a two   one-sample two-tailed t-tests were conducted, testing the null hypothesis that respectively $\mu_r^{E,1}=r^{E,1}$ and $\mu_r^{E,2}=r^{E,2}$. Both tests passed successfully with a p-value respectively equal to 0.6 and 0.1.       
Finally we computed expected values $\rho_{\infty}^{E,\nu}= \rho_0^{E,\nu}$ (Tab.3). We obtained a value  $\rho_{\infty}^{E,1}= 0.05 $ ($95\%$ c.i.: $  0.03-0.07$) for the first wave  and $\rho_{\infty}^{E,2}= 0.11 $  ($95\%$ c.i.: $ 0.08-0.14$) for the second one. Both results seem in  good agreement with the observations provided worldwide  for the incidence rate on monthly scale [ECDC] and  the deviation between the two values  accounts for the observed growing of the incidence rate  passing from the first to the second wave of infection.
 Moreover, since parameters $r^{E,1}$ and $r^{E,2}$ are very close the different values  resulting from incidence rate estimations $\rho_{\infty}^{E,\nu}$, are mainly due by parameters $\theta^{E,\nu}$, which pass from $\theta^{E,1}=34.92$ ($95\%$ c.i.: $ 21.08-57.46$) for the first wave to $\theta^{E,1}12.86$ ($95\%$ c.i.: $ 6.07-27.02$) for the second one. A decreasing $\theta^{E,\nu}$ means an increasing $\delta^{E,\nu}$, which corresponds to an increasing $d$,  the population number being constant. This could be explained by an increasing viral load, conceivably due to the inflow of a more aggressive  variant of the virus, associated with an   increasing  initial sick population $w$, since the parameter $r^{E,\nu}$ is proved to remain constant varying $\nu$.

\begin{figure}[H]
\centering
\includegraphics[width=\textwidth]{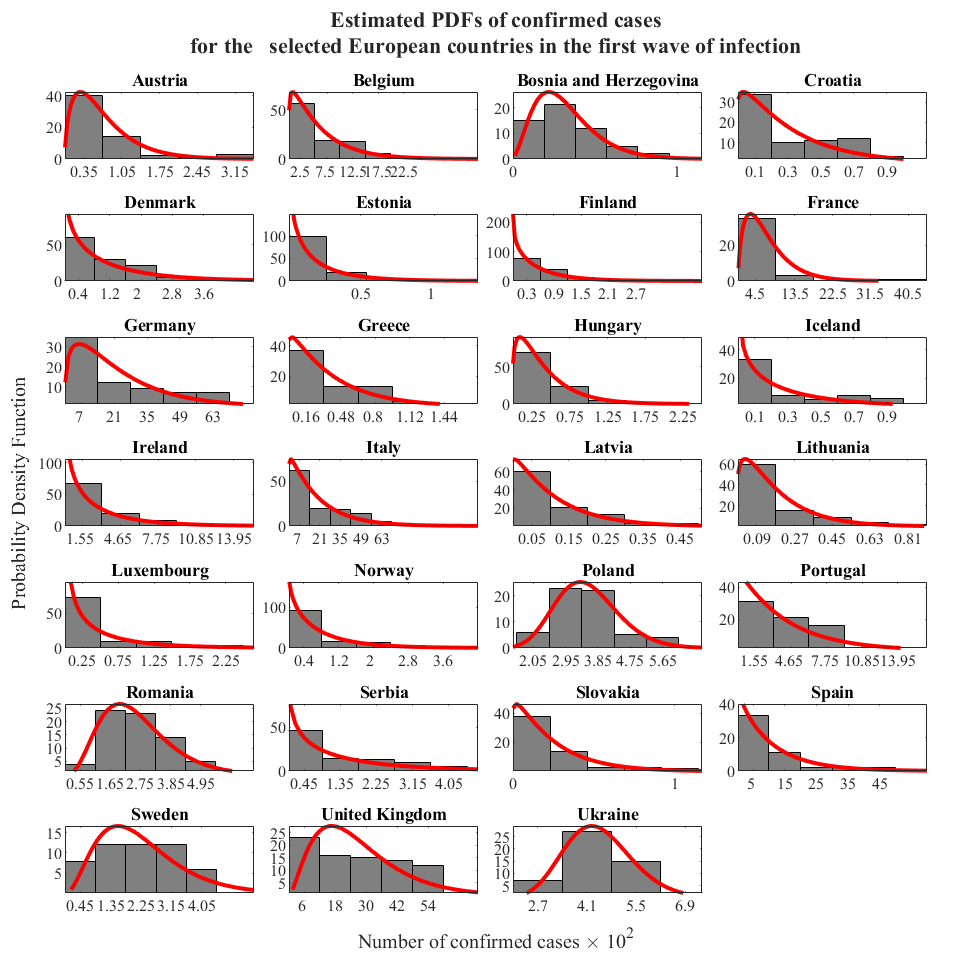}
\caption{  Negative Binomial PDFs (red line) of the confirmed cases; the group of 27 successful European countries, during the first waves of infection, was considered. Histograms of observed confirmed cases were also reported (bar plot). Black dashed lines represent upper and lower bounds, with $95\%$ c.i.}
\end{figure}

\begin{figure}[H]
\centering
\center{\includegraphics[width=\textwidth]
        {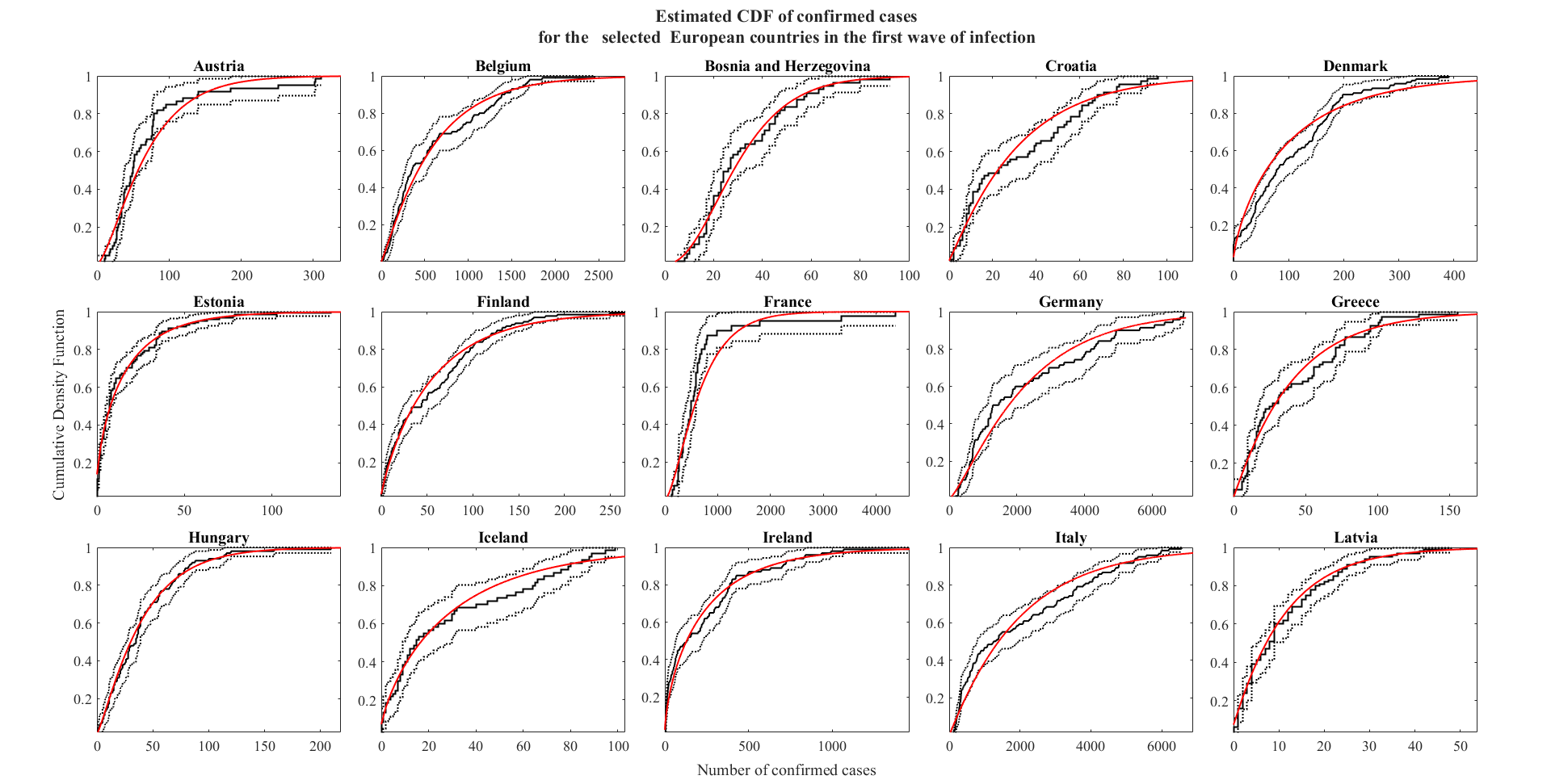}}
\caption{   Negative Binomial CDFs (red line) of the  confirmed cases; the group of 27 successful European countries, during the first waves of infection, was considered. Empirical CDFs of confirmed cases was reported (black line); black dashed lines represent upper and lower bounds, with $95\%$ c.i..}
\end{figure}

\begin{figure}[H]
\centering
\center{\includegraphics[width=\textwidth]
        {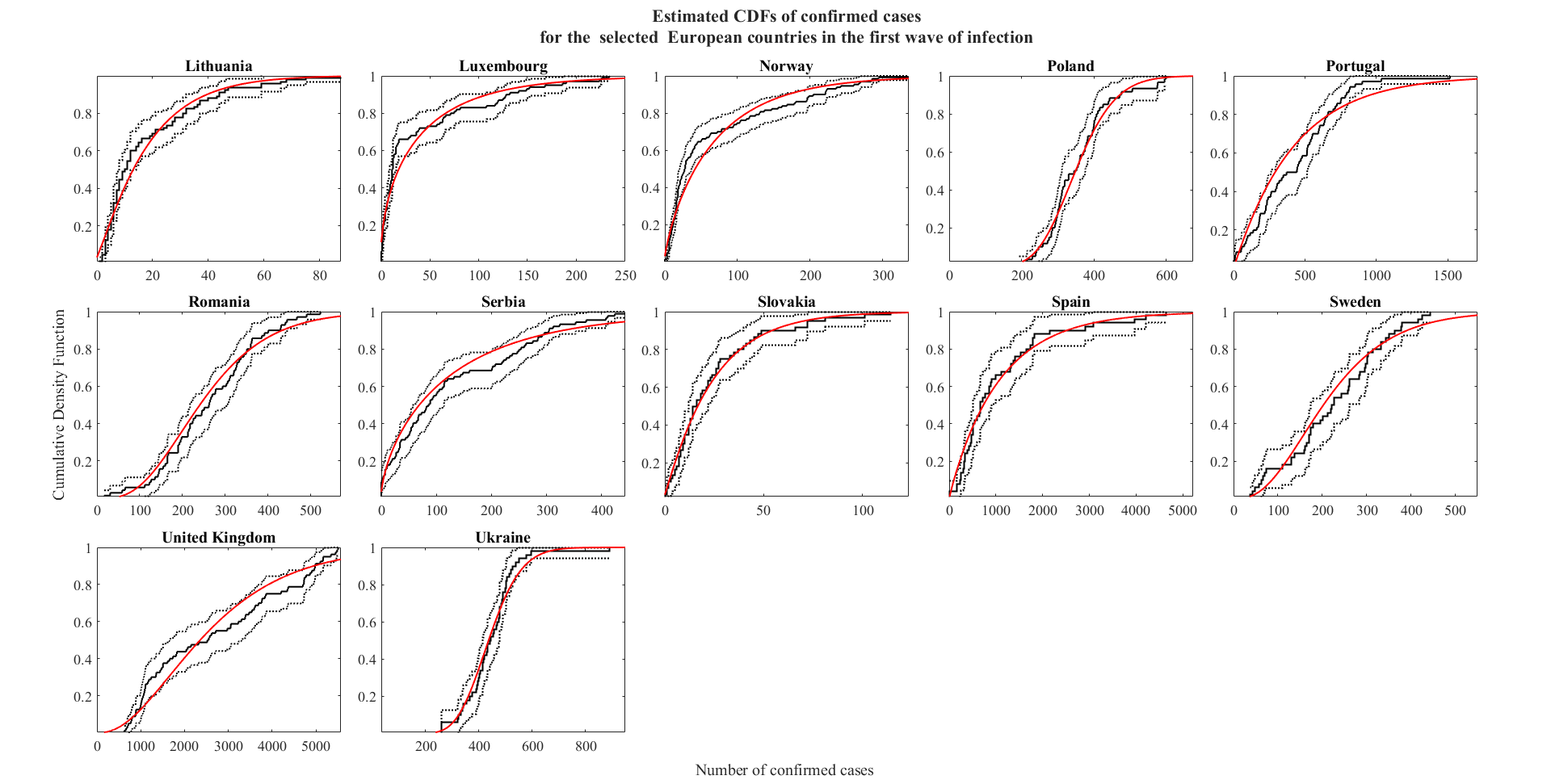}}
\caption{  Negative Binomial CDFs (red line) of the  confirmed cases; the group of 27 successful European countries, during the first waves of infection, was considered. Empirical CDFs of confirmed cases was reported (black line); black dashed lines represent upper and lower bounds, with $95\%$ c.i..}
\end{figure}
%
%
%
%
\begin{figure}[H]
\centering
\center{\includegraphics[width=\textwidth]
        {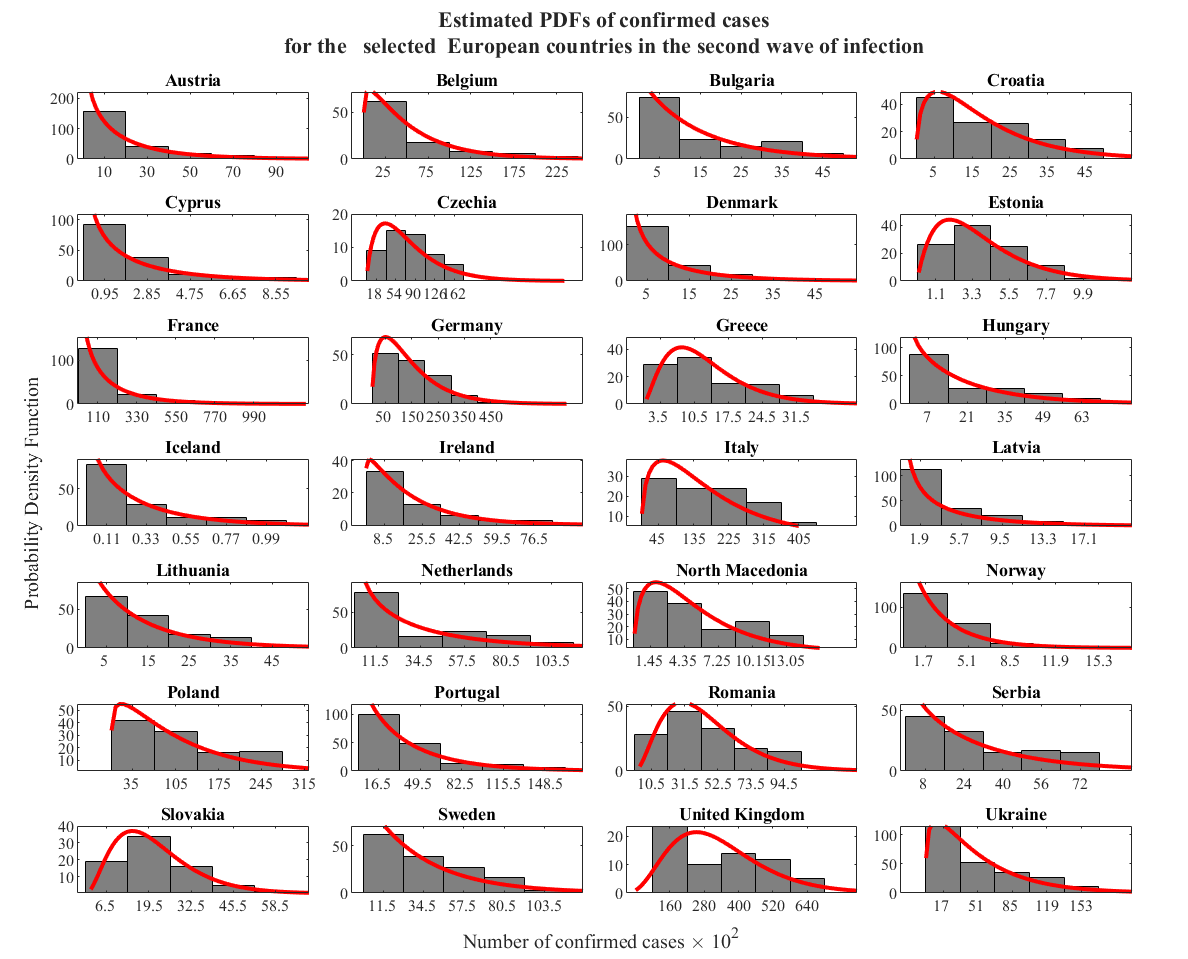}}
\caption{  Negative Binomial PDFs (red line) of the confirmed cases; the group of 28 successful European countries, during the first waves of infection, was considered. Histograms of observed confirmed cases were also reported (bar plot). Black dashed lines represent upper and lower bounds, with $95\%$ c.i..}
\end{figure}
\begin{figure}[H]
\centering
\center{\includegraphics[width=\textwidth]
        {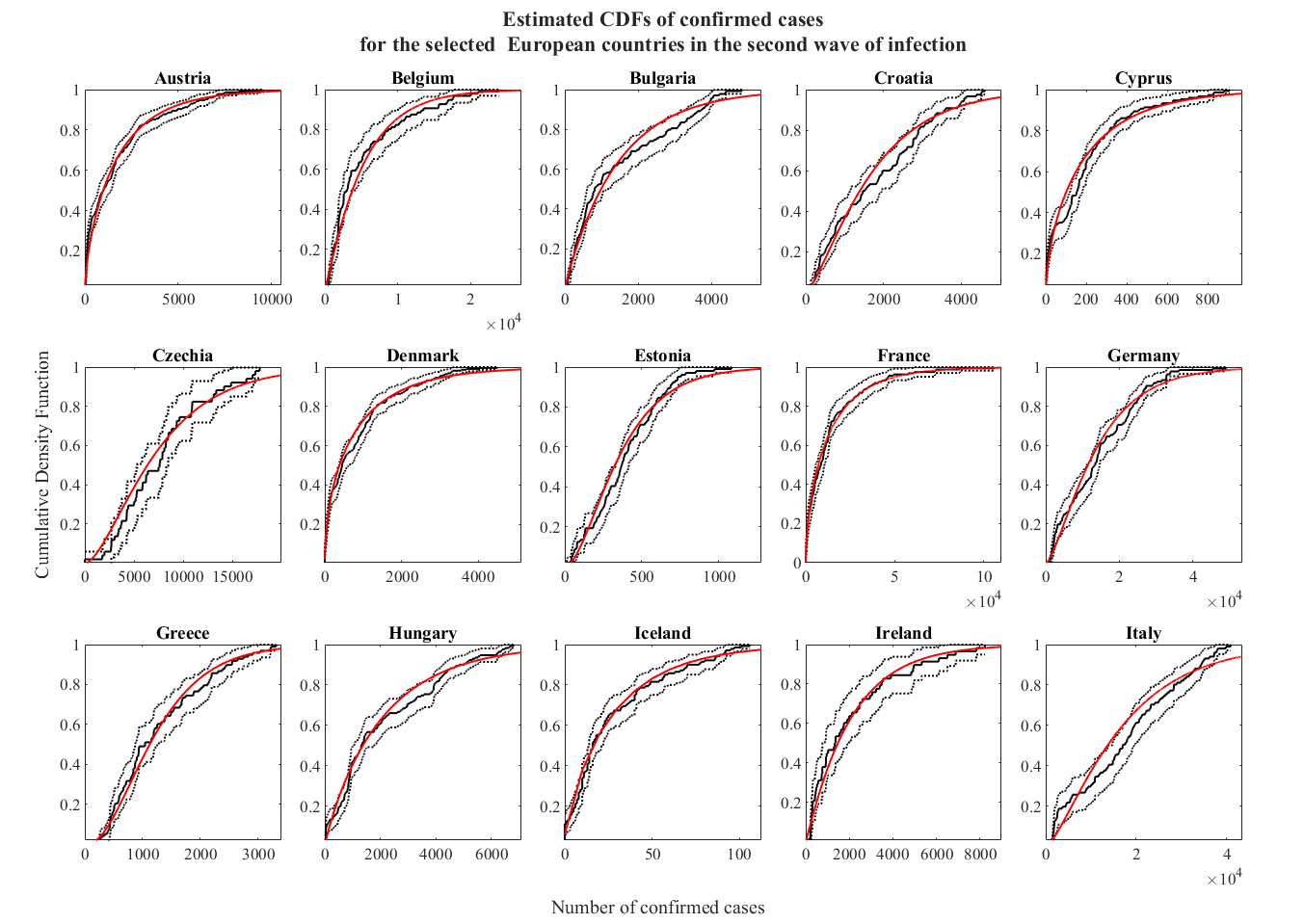}}
\caption{   Negative Binomial CDFs (red line) of the  confirmed cases; the group of 28 successful European countries, during the first waves of infection, was considered. Empirical CDFs of confirmed cases was reported (black line); black dashed lines represent upper and lower bounds, with $95\%$ c.i..}
\end{figure}

\begin{figure}[H]
\centering
\center{\includegraphics[width=\textwidth]
        {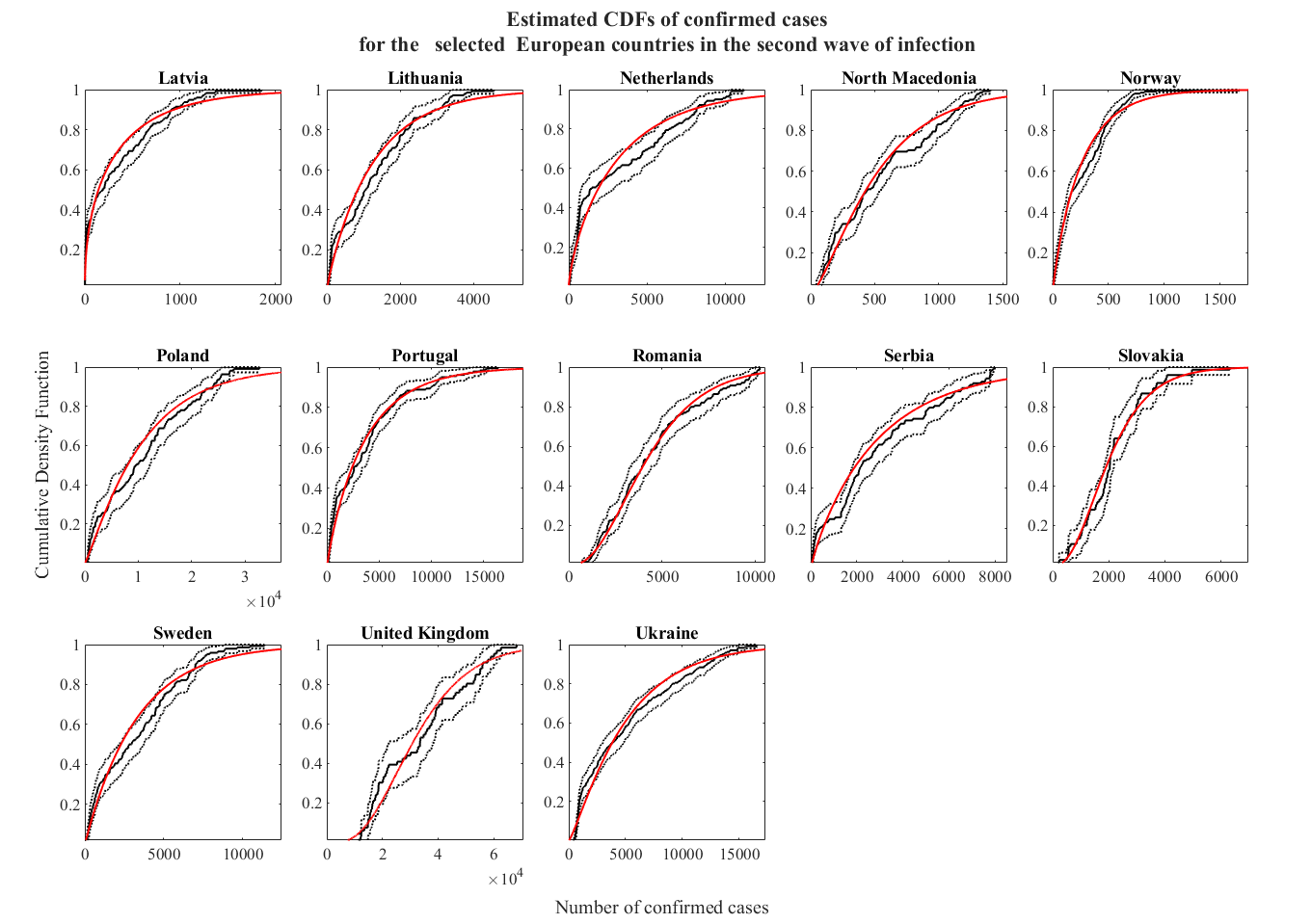}}
\caption{  Negative Binomial CDFs (red line) of the  confirmed cases; the group of 28 successful European countries, during the first waves of infection, was considered. Empirical CDFs of confirmed cases was reported (black line); black dashed lines represent upper and lower bounds, with $95\%$ c.i..}
\end{figure}
\begin{figure}[H]
\centering
\center{\includegraphics[width=\textwidth]
        {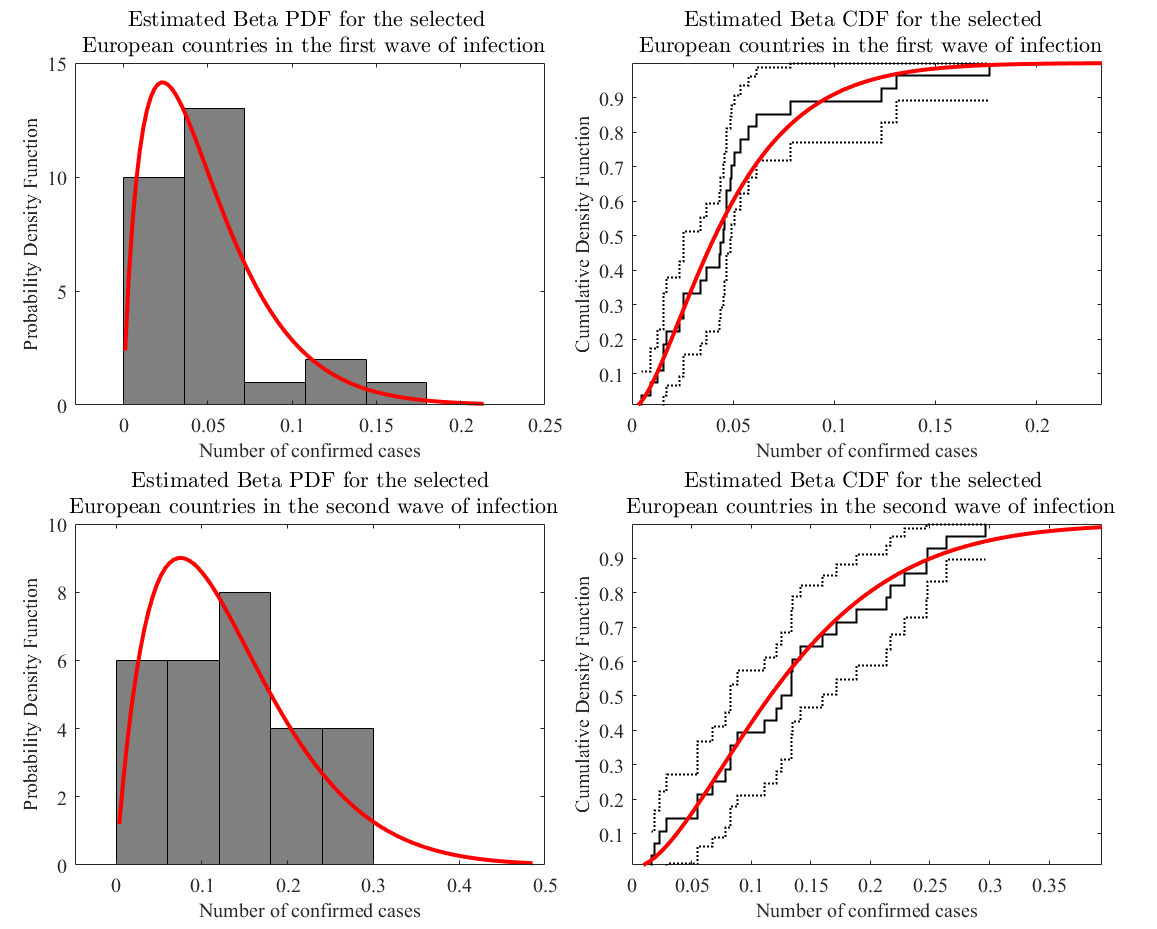}}
\caption{ On the right:  Beta PDFs of the ratio $R_{i,\infty}^{E,\nu}$ (red line) with relative histogram (bar plot) during the first (up) and second (down) waves of infection. On the left:  CDFs  corresponding to the above PDFs (red line); black dashed lines represent upper  and lower  bound, with $95\%$ c.i..}
\end{figure}

\begin{table}[H]
\centering
\begin{tabular}{l r r r }
\hline
\hline
& & & \\
\textbf{Countries} & $\bm{r_i^{E,1} \pm 95\% \textbf{c.i.}}$ & $\bm{p_i^{E,1}  \pm 95\% \textbf{c.i.}} \; \bm{[10^{-2}]}$ & $\bm{p-\textbf{value}}$ \\
& & & \\
\hline
Austria  &	1.70 $\pm$	0.57 &	2.33 $\pm$	0.92 &	0.15  \\
Belgium  &	1.11 $\pm$	0.28  &	0.19 $\pm$	0.06  &	0.54  \\
Bos.-Herzeg.  &	3.29 $\pm$	1.31  &	9.33 $\pm$	3.62  &	0.72  \\
Croatia  &	1.12 $\pm$	0.36  &	3.44 $\pm$	1.30 &	0.24 \\
Denmark  &	0.70 $\pm$	0.17  &	0.67 $\pm$	0.22  &	0.02  \\
Estonia  &	0.59 $\pm$	0.14  &	3.59 $\pm$	1.17  &	0.01  \\
Finland  &	0.83 $\pm$	0.19  &	1.48 $\pm$	0.43  &	0.15  \\
France  &	1.73 $\pm$	0.70  &	0.25 $\pm$	0.12  &	0.06 \\
Germany  &	1.35 $\pm$	0.40  &	0.06 $\pm$	0.03  &	0.22  \\
Greece  &	1.06 $\pm$	0.35  &	2.57 $\pm$	1.02  &	0.49  \\
Hungary  &	1.30 $\pm$	0.35  &	3.12 $\pm$	0.97  &	0.70  \\
Iceland  &	0.69 $\pm$	0.24  &	2.34 $\pm$	1.04  &	0.66  \\
Ireland  &	0.62 $\pm$	0.14  &	0.26 $\pm$	0.09  &	0.38  \\
Italy  &	1.04 $\pm$	0.23 &	0.06 $\pm$	0.02  &	0.12  \\
Latvia  &	1.08 $\pm$	0.32  &	8.98 $\pm$	2.94  &	0.44  \\
Lithuania  & 1.27 $\pm$	0.36  &	6.55 $\pm$	2.11  &	0.08  \\
Luxembourg  &	0.50 $\pm$	0.12  &	1.24 $\pm$	0.46  &	0.01  \\
Norway  &	0.77 $\pm$	0.16  &	1.11 $\pm$	0.33  &	0.01  \\
Poland  &	17.57 $\pm$	6.53  &	4.74 $\pm$	1.69  &	0.66  \\
Portugal  &	1.02 $\pm$	0.32  &	0.25 $\pm$	0.10  &	0.06 \\
Romania  &	4.20 $\pm$	1.38  &	1.59 $\pm$	0.55  &	0.71  \\
Serbia  &	0.62 $\pm$	0.17  &	0.49 $\pm$	0.19  &	0.29  \\
Slovakia  &	1.10 $\pm$	0.39  &	4.41 $\pm$	1.82  &	0.87  \\
Spain  &	0.94 $\pm$	0.34  &	0.09 $\pm$	0.05  &	0.32  \\
Sweden  &	3.32 $\pm$	1.27  &	1.46 $\pm$	0.59  &	0.56  \\
Un.Kingdom &	2.57 $\pm$	0.75  &	0.15 $\pm$	0.04  &	0.23  \\
Ukraine  &	22.04 $\pm$	9.01  &	4.73 $\pm$	1.87  &	0.73 \\

\hline
\end{tabular}
\caption{Fitting parameters $r_i^{E,1}$ and $p_i^{E,1}$  of the Negative Binomial PDFs for the selected European countries, in the first wave of infection.  Best p-values resulting from both KS and $\chi^2$ test are also reported.  }
\end{table}  
\begin{table}[H]
\centering
\begin{tabular}{l r r r }
\hline
\hline
& & & \\
\textbf{Countries} & $\bm{r_i^{E,2} \pm 95\% \textbf{c.i.}}$ & $\bm{p_i^{E,2} \pm 95\% \textbf{c.i.}}\;[10^{-2}]$ & $\bm{p-\textbf{value}}$ \\
& & & \\
\hline
Austria  &	0.67   $\pm$	0.10  &	0.04 $\pm$	0.01  &	0.04  \\
Belgium  &	1.13  $\pm$	0.28  &	0.03 $\pm$	0.01  &	0.14  \\
Bulgaria &	0.97  $\pm$	0.20  &	0.07 $\pm$	0.02  &	0.30 \\
Croatia &	1.48  $\pm$	0.34  &	0.09 $\pm$	0.03  &	0.12  \\
Cyprus &	0.59  $\pm$	0.12 &	0.31 $\pm$	0.09  &	0.01  \\
Czechia &	1.79 $\pm$	0.67 &	0.03 $\pm$	0.01  &	0.51  \\
Denmark &	0.51  $\pm$	0.08  &	0.07 $\pm$	0.02  &	0.12 \\
Estonia &	2.03  $\pm$	0.52  &	0.53 $\pm$	0.16  &	0.13  \\
France &	0.58 $\pm$	0.10  &	0.01 $\pm$	0.01  &	0.77  \\
Germany &	1.54 $\pm$	0.33  &	0.02 $\pm$	0.01  &	0.14  \\
Greece &	2.57  $\pm$	0.68  &	0.20 $\pm$	0.06  &	0.34  \\
Hungary &	0.88  $\pm$	0.16 &	0.05 $\pm$	0.02  &	0.18 \\
Iceland &	0.77  $\pm$	0.18  &	2.77 $\pm$	0.82  &	0.15  \\
Ireland &	1.06  $\pm$	0.34 &	0.06 $\pm$	0.03  &	0.70  \\
Italy &	1.47  $\pm$	0.37  &	0.01 $\pm$	0.01  &	0.05  \\
Latvia &	0.46 $\pm$	0.08  &	0.14 $\pm$	0.04  &	0.04  \\
Lithuania &	0.90  $\pm$	0.18  &	0.08 $\pm$	0.02  &	0.10  \\
Netherlands &	0.72 $\pm$	0.14  &	0.03 $\pm$	0.01  &	0.02  \\
Nor.Maced. &	1.53  $\pm$	0.32 &	0.29 $\pm$	0.08  &	0.11 \\
Norway &	0.87 $\pm$	0.15  &	0.34 $\pm$	0.08  &	0.04  \\
Poland &	1.17  $\pm$	0.27  &	0.02 $\pm$	0.01  &	0.18  \\
Portugal &	0.87  $\pm$	0.15  &	0.03 $\pm$	0.01  &	0.06  \\
Romania &	3.02  $\pm$	0.67  &	0.07 $\pm$	0.02  &	0.90 \\
Serbia &	0.87  $\pm$	0.19  &	0.04 $\pm$	0.01  &	0.01  \\
Slovakia &	3.04  $\pm$	0.92  &	0.15 $\pm$	0.05  &	0.33  \\
Sweden &	1.01  $\pm$	0.20  &	0.04 $\pm$	0.01  &	0.05 \\
Un.Kingdom &	4.13  $\pm$	1.35  &	0.02 $\pm$	0.01  &	0.27 \\
Ukraine &	1.25  $\pm$	0.20 &	0.03 $\pm$	0.01  &	0.02  \\

\hline
\end{tabular}
\caption{Fitting parameters $r_i^{E,2}$ and $p_i^{E,2}$  of the Negative Binomial PDFs for the selected European countries, in the second wave of infection.  Best p-values resulting from both KS and $\chi^2$ test are also reported.}
\end{table}

\begin{table}[H]
\small
\centering
\begin{tabular}{l l l l l l l  }
\hline
& \vline & & &   \\
\textbf{Waves of} & \vline & & &   \\
 \textbf{infection} & \vline & & &   \\ 
 \textbf{(W.I-II)} &  \vline & $\bm{r^{E,\nu} \; (95\% \textbf{c.i.})}$ & $\bm{\theta^{E,\nu} \;(95\% \textbf{c.i.})}$  &$\bm{\rho_{\infty}^{E,\nu}\; (95\% \textbf{c.i.})}$ & $\mu_r^{E,\nu} \; (95\%$ c.i.)&$\mu_p^{E,\nu} \; (95\%$ c.i.)$[ 10^{-2}]$ \\
\hline
\hline
&  \vline & & &  \\
W.I &\vline &	$1.80 \;  (1.07-3.02)$  &	 34.92 $(21.08-57.46)$   & $0.05 \;(0.03-0.07) $&  	$1.48$ ( $1.34 - 1.62 $) & $3.17 $  ($2.06  - 4.28  $)\\
& \vline & & &   \\
\hline
& \vline & & &   \\
W.II &\vline &	$1.95 \;(1.10-3.46)$ &	$12.86 \; (6.07-27.02)$   & $0.11 \;(0.08-0.14)$ &  	 1.68 ( $1.54-1.82 $)  & 0.04   ($0.03  - 0.05  $) \\
& \vline & & &   \\

\hline
\end{tabular}
\caption{Fitting parameters $r^{E,\nu}$ and $\theta^{E,\nu}$ of the Beta PDFs for the first and second waves of infection. Mean values $\rho_{\infty}^{E,\nu}$ and p-values, resulting from both KS and $\chi^2$ test, are also reported.}
\end{table}  

\subsection{Italy }
The entire procedure described above  for the European case was repeated for  the  21 Italian regions, in particular the sequence of statistical tests with the same significance levels. Fig.8-11 show 14 and 13 successful Negative Binomials PDFs and CDFs,  corresponding to the selected regions during respectively the first and  the second wave of infection. Tab.4-5 report the estimated fitting parameters,  for both the two waves.
Sample means and variance of  $r^{I,\nu}$ and $p^{I,\nu}$ were 
estimated obtaining the following results:  $\mu_r^{I,1}=0.52$ $95\%$ c.i. $(0.39 - 0.65)$, $\mu_p^{I,1}=1.11$ $95\%$ c.i. $(0.81 - 1.41)$, $\mu_r^{I,2}=0.45$ $95\%$ c.i. $(0.37 - 0.53)$ and $\mu_p^{I,2}=0.12$ $95\%$ c.i. $(0.07 - 0.17)$. An analogue sequence of one-sample two-tailed t-tests on both the variances were successfully performed. Two one-sample two-tailed t-tests for comparing $\mu_*^{I,1}$ and $\mu_*^{I,2}$ were also successfully performed. Results seem internally consistent and consistent with the European scenario. However, parameters $r^{I,\nu}$ and $\theta^{I,\nu}$,  estimated by the $B\big(r^{I,\nu},\theta^{I,\nu}\big)$, show some deviation from European case that need to be discussed. In fact, while $r^{I,1}$ value can be still considered in agreement with previous results, $r^{I,2}$ is clearly far from all respective value changing $g$ or $\nu$.     
Nevertheless, a look to the Tab.6 shows that values of $\rho_{\infty}^{I,1}=0.04$ $95\%$ c.i. $(0.02 - 0.06)$ and   $\rho_{\infty}^{I,2}=0.07$ $95\%$ c.i. $(0.05 - 0.09)$ seems in agreement respectively with the first and second waves related to the European case. 
Actually the reason why the strong differences between values of $r^{I,2}$ estimated from the corresponding Beta and the mean values of  $r_i^{I,2}$ obtained from the single regions, must be sought in that a small amount of data are available. Moreover, similarly to the European case (Fig.15), in the second wave the ratio $R_{i,n}^{I,2}$ have not reached yet stable values. Crossed t-tests, as described in the European case were also performed comparing all corresponding Italian and European estimated parameters with successful  results.

 \begin{figure}[H]
\centering
\center{\includegraphics[width=\textwidth]
        {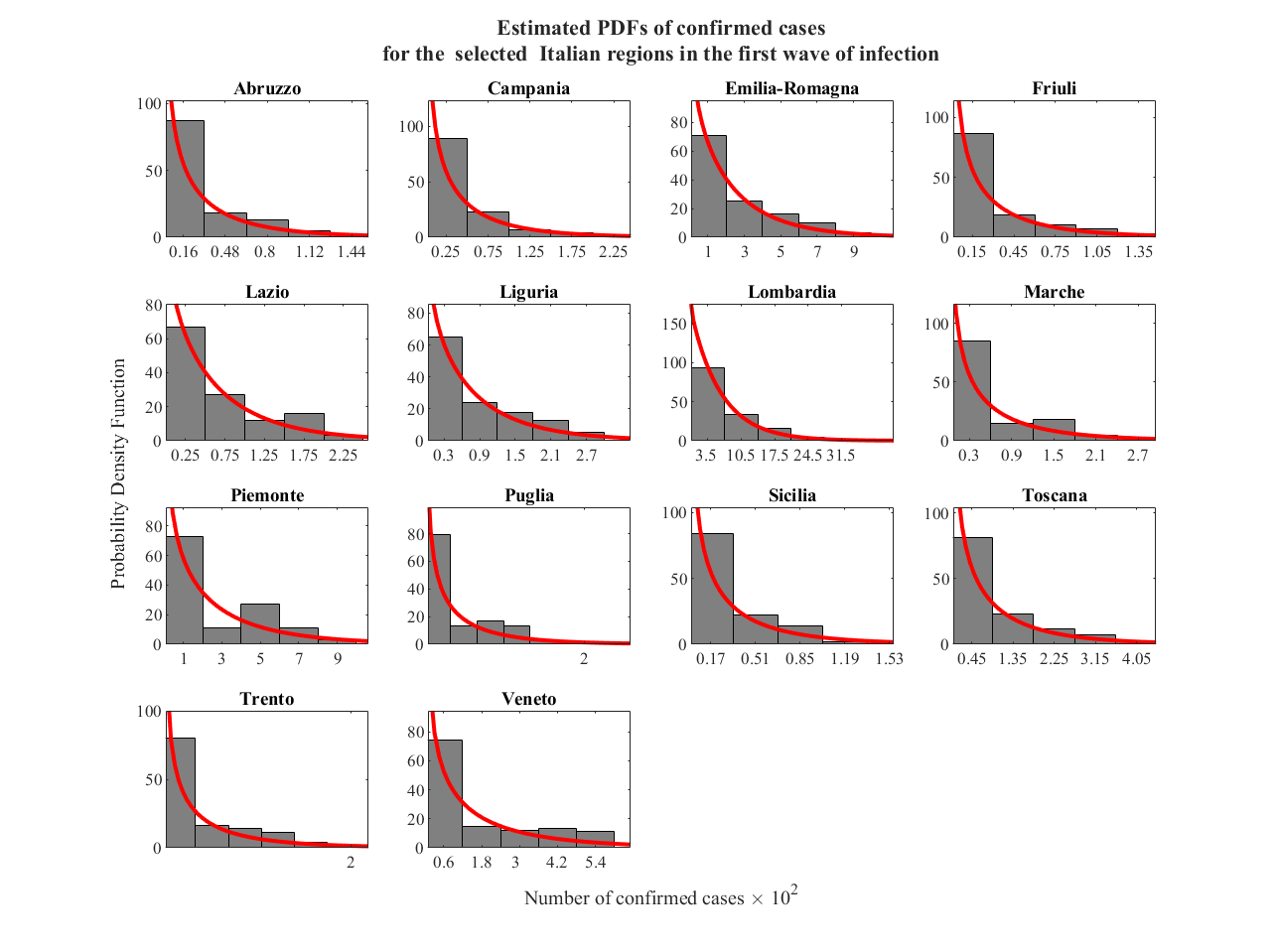}}
\caption{  Negative Binomial PDFs (red line) of the confirmed cases; the group of 14 successful Italian regions, during the first waves of infection, was considered. Histograms of observed confirmed cases were also reported (bar plot). Black dashed lines represent upper and lower bounds, with $95\%$ c.i..}
\end{figure}

\begin{figure}[H]
\centering
\center{\includegraphics[width=\textwidth]
        {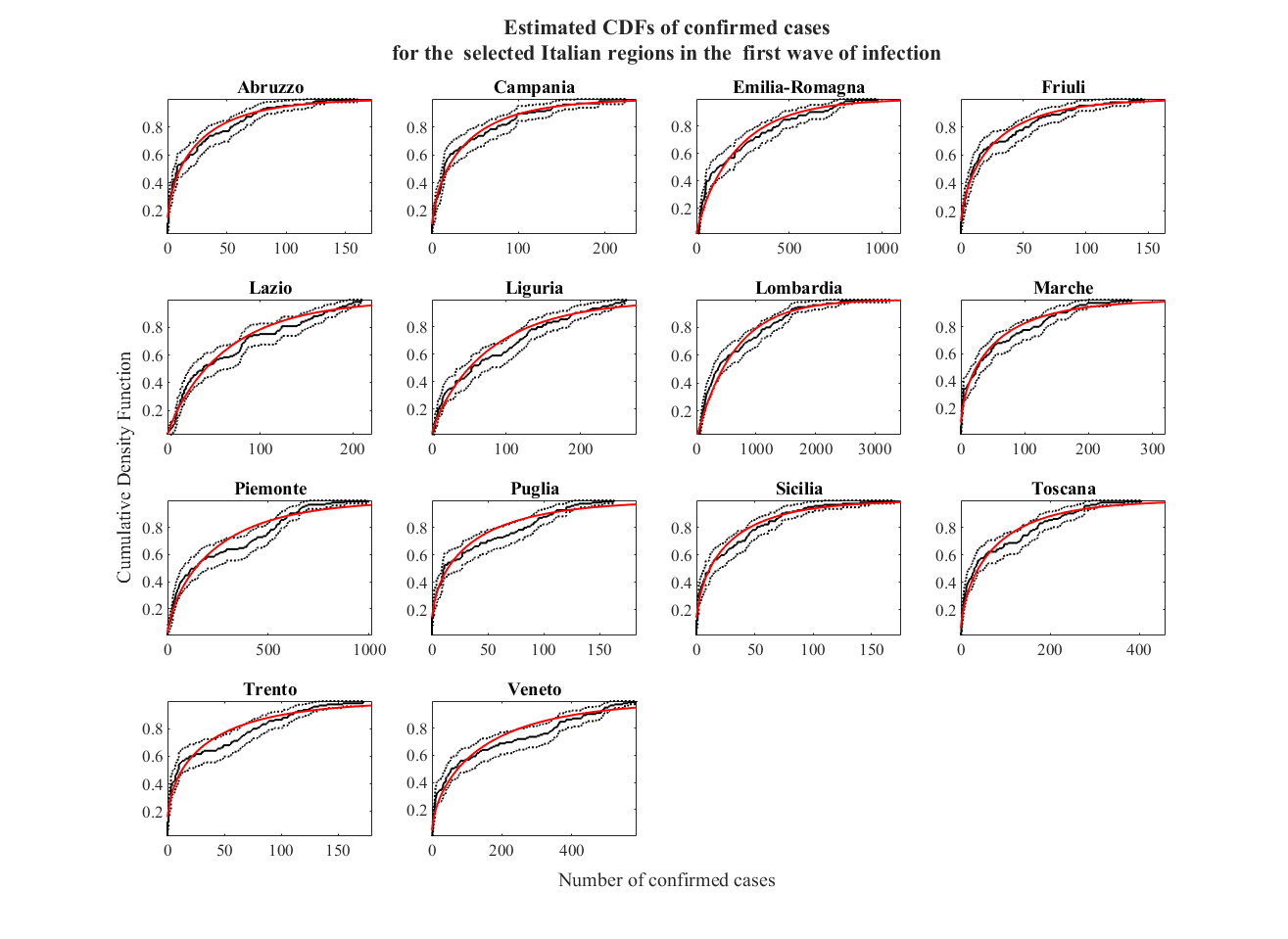}}
\caption{  Negative Binomial CDFs (red line) of the  confirmed cases; the group of 14 successful Italian regions, during the first waves of infection, was considered. Empirical CDFs of confirmed cases was reported (black line); black dashed lines represent upper and lower bounds, with $95\%$ c.i..}
\end{figure}
%
%
%
%
\begin{figure}[H]
\centering
\center{\includegraphics[width=\textwidth]
        {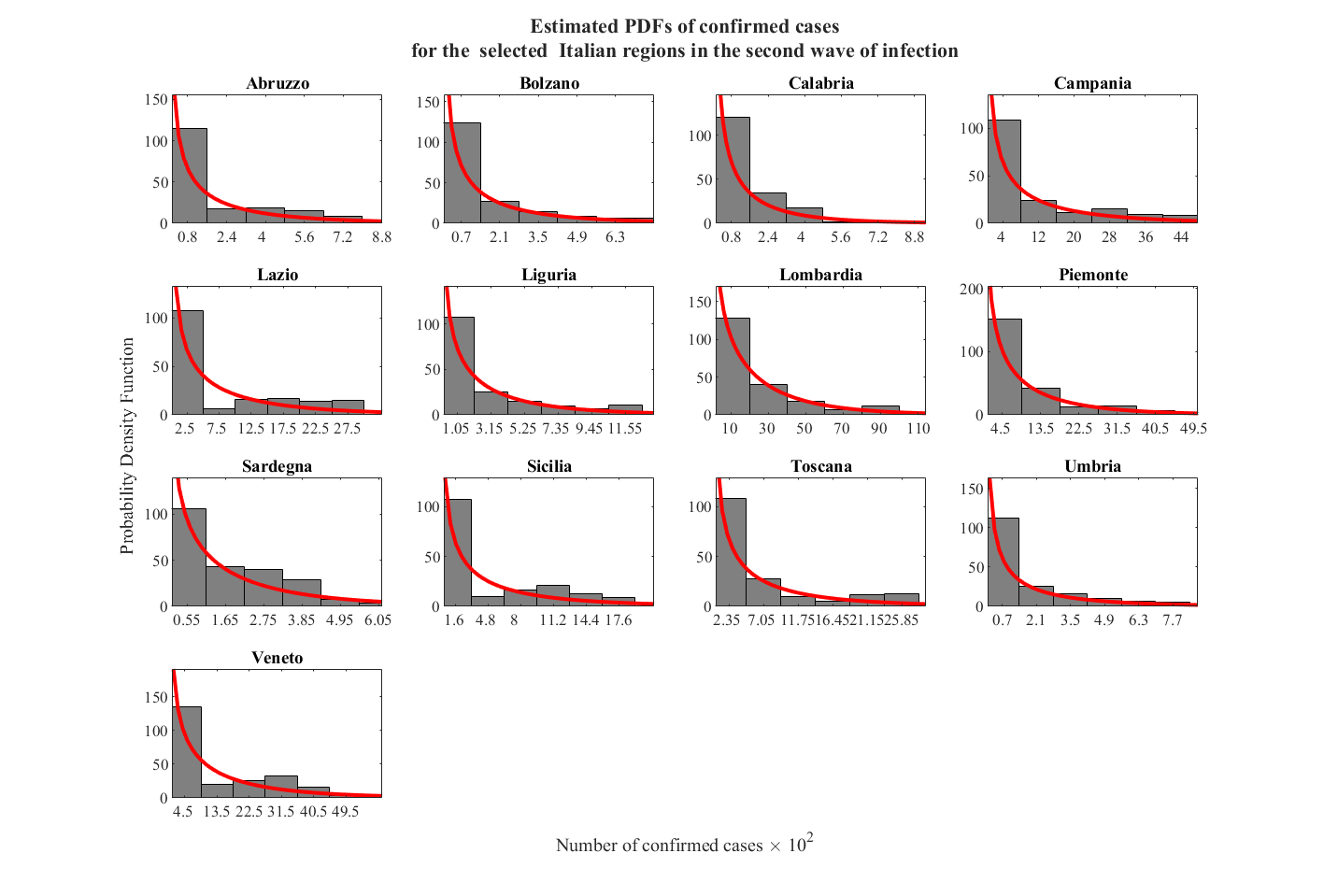}}
\caption{  Negative Binomial PDFs (red line) of the confirmed cases; the group of 13 successful Italian regions, during the first waves of infection, was considered. Histograms of observed confirmed cases were also reported (bar plot). Black dashed lines represent upper and lower bounds, with $95\%$ c.i..}
\end{figure}
\begin{figure}[H]
\centering
\center{\includegraphics[width=\textwidth]
        {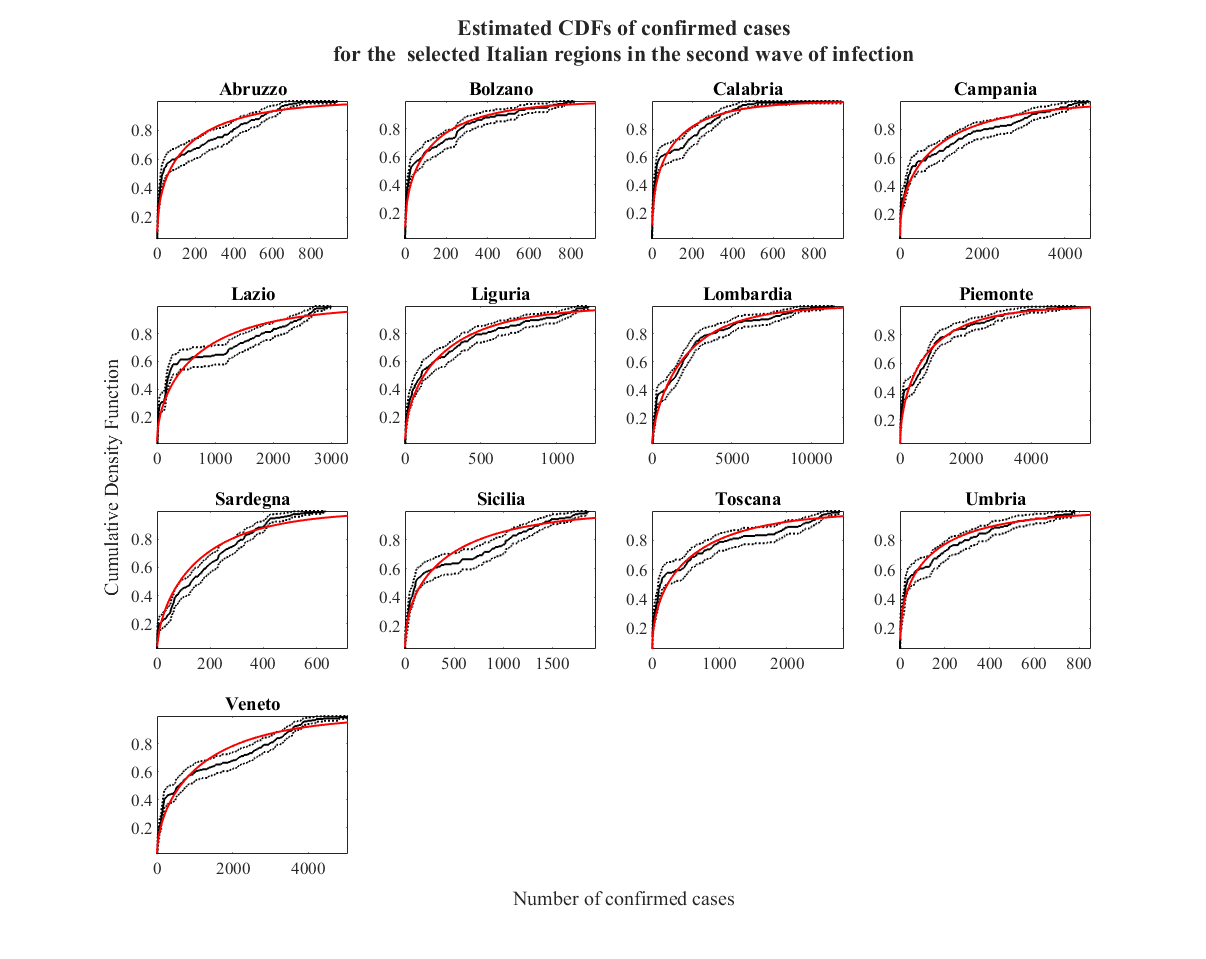}}
\caption{  Negative Binomial CDFs (red line) of the  confirmed cases; the group of 13 successful Italian regions, during the first waves of infection, was considered. Empirical CDFs of confirmed cases was reported (black line); black dashed lines represent upper and lower bounds, with $95\%$ c.i..}
\end{figure}
\begin{figure}[H]
\centering
\center{\includegraphics[width=\textwidth]
        {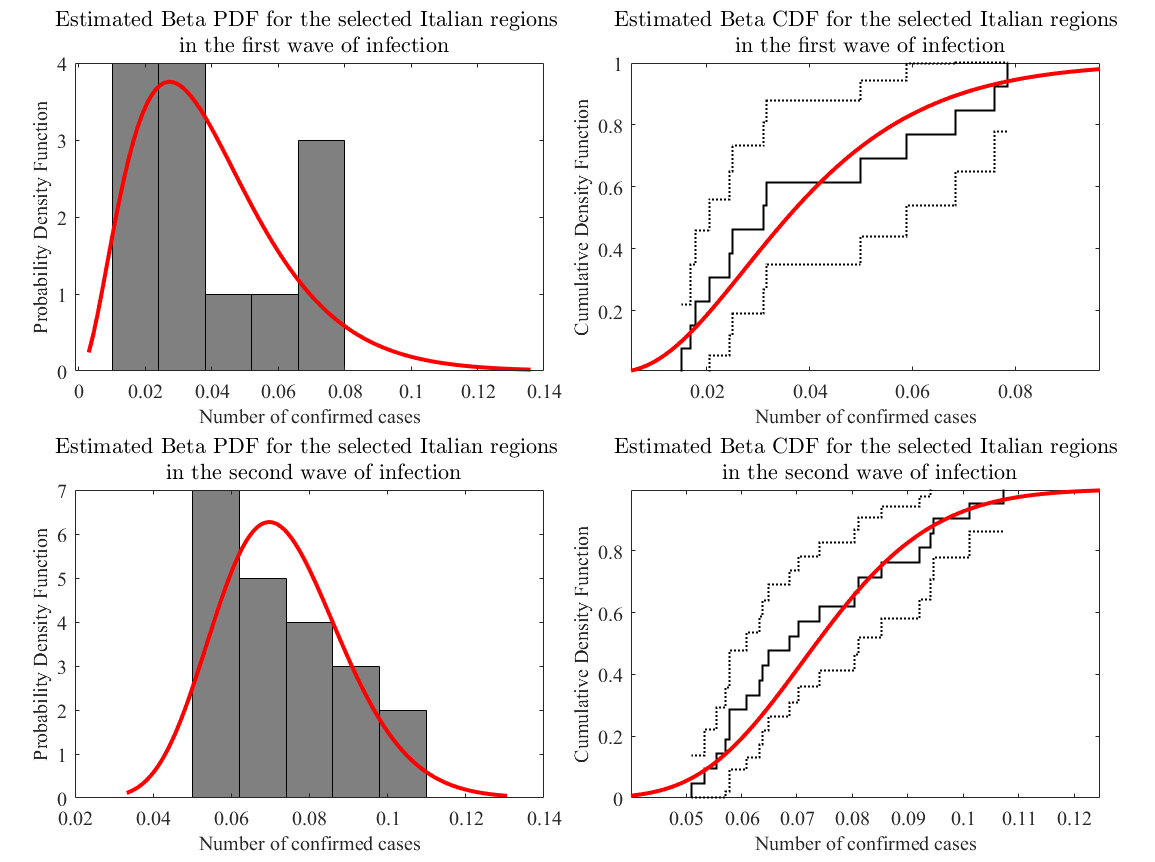}}
\caption{ On the right:  Beta PDFs of the ratio $R_{i,\infty}^{I,\nu}$ (red line) with relative histogram (bar plot) during the first (up) and second (down) waves of infection. On the left:  CDFs  corresponding to the above PDFs (red line); black dashed lines represent upper  and lower  bound, with $95\%$ c.i..}
\end{figure}

\begin{table}[H]
\centering
\begin{tabular}{l r r r }
\hline
\hline
& & & \\
\textbf{Countries} & $\bm{r_i^{I,1} \pm 95\% \textbf{c.i.}}$ & $\bm{p_i^{I,1} \pm 95\% \textbf{c.i.}\;[10^{-2}]}$ & $\bm{p-\textbf{value}}$ \\
& & & \\
\hline
Abruzzo  &	0.48 $\pm$	0.11   &	1.78 $\pm$	0.61   &	0.01  \\
Campania   &	0.52 $\pm$	0.12   &	1.32 $\pm$	0.44   &	0.12  \\
Emilia-Romagna & 0.82 $\pm$	0.18   &	0.37 $\pm$ 0.11  & 0.02  \\
Friuli   &	0.51 $\pm$	0.12   &	1.89 $\pm$	0.64   &	0.03  \\
Lazio   &	0.84 $\pm$	0.19   &	1.29 $\pm$	0.39   &	0.29  \\
Liguria   &	0.81 $\pm$	0.18   &	1.01 $\pm$	0.30   &	0.11  \\
Lombardia   &0.99 $\pm$	0.20   &	0.16 $\pm$	0.04   &	0.19  \\
Marche   &	0.50 $\pm$	0.11   &	0.91 $\pm$	0.31   &	0.14  \\
Piemonte   &	0.65 $\pm$	0.14   &	0.26 $\pm$	0.08   &	0.03  \\
Puglia   &	0.44 $\pm$	0.10   &	1.21 $\pm$	0.43   &	0.02  \\
Sicilia   &	0.47 $\pm$	0.11   &	1.68 $\pm$	0.58   &	0.01  \\
Toscana   &	0.54 $\pm$	0.12   &	0.66 $\pm$	0.21   &	0.16  \\
Trento   &	0.41 $\pm$	0.10   &	1.14 $ \pm$	0.41   &	0.01  \\
Veneto   &	0.54 $\pm$	0.11   &	0.35 $ \pm$	0.12   &	0.13  \\
\hline
\end{tabular}
\caption{Fitting parameters $r_i^{I,1}$ and $p_i^{I,1}$  of the Negative Binomial PDFs for the selected Italian regions, in the second wave of infection.  Best p-values resulting from both KS and $\chi^2$ test are also reported.}
\end{table}

\begin{table}[H]
\centering
\begin{tabular}{l r r r }
\hline
\hline
& & & \\
\textbf{Countries} & $\bm{r_i^{I,2} \pm 95\% \textbf{c.i.}}$ & $\bm{p_i^{I,2} \pm 95\% \textbf{c.i.}\;[10^{-2}]}$ & $\bm{p-\textbf{value}}$ \\
& & & \\
\hline
Abruzzo   &	0.39 $ \pm$	0.07   &	0.23 $ \pm$	0.07   &	0.02  \\
Bolzano   &	0.40 $ \pm$	0.07   &	0.28 $\pm$	0.08   &	0.01  \\
Calabria   &	0.38 $ \pm$	0.07   &	0.34 $\pm$	0.10   &	0.02  \\
Campania   &	0.40 $\pm$	0.07   &	0.04 $\pm$	0.02   &	0.19  \\
Lazio   &	0.50 $\pm$	0.09   &	0.07 $\pm$	0.02   &	0.01  \\
Liguria   &	0.51 $\pm$	0.09   &	0.19 $\pm$	0.06   &	0.36  \\
Lombardia  &	0.66 $\pm$	0.11   &	0.03 $\pm$	0.01   &	0.01 \\
Piemonte   &	0.49 $\pm$	0.07   &	0.06 $\pm$	0.02   &	0.02  \\
Sardegna   &	0.57 $\pm$	0.10   &	0.34 $\pm$	0.09   &	0.01  \\
Sicilia   &	0.41 $\pm$	0.07   &	0.09 $\pm$	0.03   &	0.02  \\
Toscana   &	0.42 $\pm$	0.07   &	0.08 $\pm$	0.03   &	0.02  \\
Umbria   &	0.35 $\pm$	0.07   &	0.24 $\pm$	0.08   &	0.01  \\
Veneto   &	0.53 $\pm$	0.08   &	0.05 $\pm$	0.01   &	0.01 \\

\hline
\end{tabular}
\caption{Fitting parameters $r_i^{I,2}$ and $p_i^{I,2}$  of the Negative Binomial PDFs for the selected Italian regions, in the second wave of infection.  Best p-values resulting from both KS and $\chi^2$ test are also reported.}
\end{table}

\begin{table}[H]
\centering
\begin{tabular}{l l l l l l }
\hline
& \vline & & & &  \\
\textbf{Waves of infection (W.I-II)} &  \vline & $\bm{r^{I,\nu}  \; (95\% \textbf{c.i.})}$ & $\bm{\theta^{I,\nu}  \; (95\% \textbf{c.i.})}$  &$\bm{\rho_{\infty}^{I,\nu} \;(95\% \textbf{c.i.})}$ & $\bm{p-\textbf{value}}$\\
\hline
\hline
&  \vline & & & & \\
W.I &\vline &	$2.85 \; (1.01-6.10)  $  & 	$68.13 \; (30.31-112.02)$   & $0.04 \; (0.02-0.06)$ &   	0.82   \\
& \vline & & & &  \\
\hline
& \vline & & & &  \\
W.II &\vline &	$21.81 \; (11.32-39.12)$ & $273.82 \;(115.04-401.10)$   & $ 0.07 \; (0.05-0.09) $ &    	0.98   \\
& \vline & & & &  \\

\hline
\end{tabular}
\caption{Fitting parameters $r^{I,\nu}$ and $\theta^{I,\nu}$ of the Beta PDFs for the first and second waves of infection. Mean values $\rho_{\infty}^{I,\nu}$ and p-values, resulting from both KS and $\chi^2$ test, are also reported.}
\end{table}  

\section{Conclusion}

Guided by heuristic evidence of some universality property, here we propose to describe the spread of (SARS-CoV-2)-infected pneumonia (COVID-19) within a probabilistic Polya urn scheme. Under general homogeneity assumptions on initial conditions and applying a multiple waves approach, we analysed European data reported on confirmed cases and diagnostic test performed. A comparative analysis at regional and national scales was performed showing  the presence of distinctive features
according to the same underlying process at different scales.
general characteristics can be extracted.  Specific patterns and key indicators slightly depend from social or geographical conditions. On the other hand some parameter seem to hold universality properties, properly identifying COVID-19 infection.
A sequence of statistical tests were performed to prove our hypothesis. 
Based on test results, for each wave of infection, we were able to consider data from each European country, or Italian
region, as $i$ different sequences of trials of a process with different population but characterized by
the same distribution for its sample average. 
This allows us to  estimate the incidence rate $\rho_{\infty}$ by the asymptotic mean of the sample average of the process. Resulting estimation of $\rho_{\infty}^{g,\nu}$, related to the  first and second wave of infection for the Italian case are  broadly in line with European one and in agreement with  real observations \cite{ECD}.

\begin{figure}[H]
\centering
\center{\includegraphics[width=\textwidth]
        {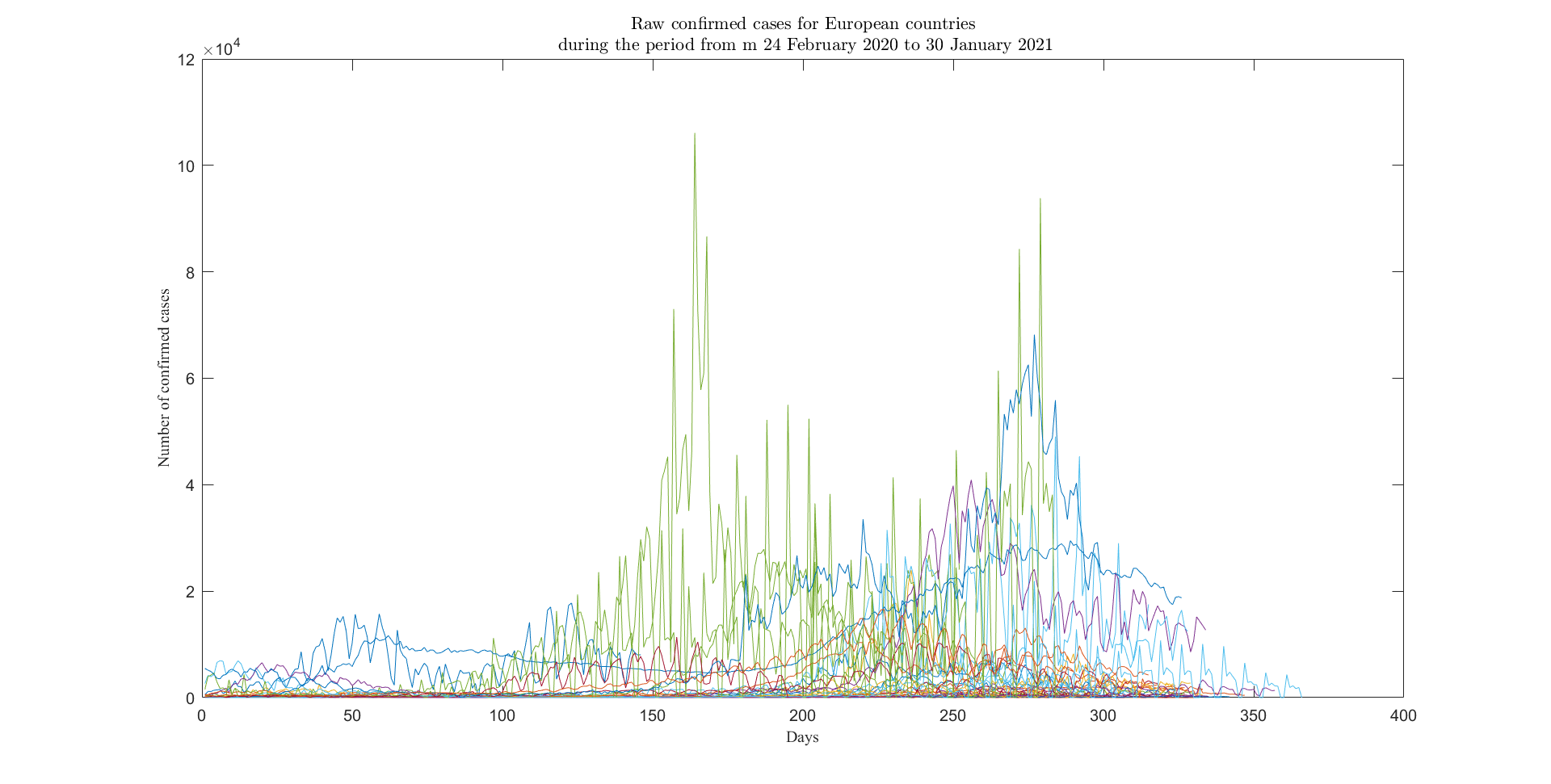}}
\caption{ Confirmed cases: row data }
\end{figure}

\begin{figure}[H]
\centering
\center{\includegraphics[width=\textwidth]
        {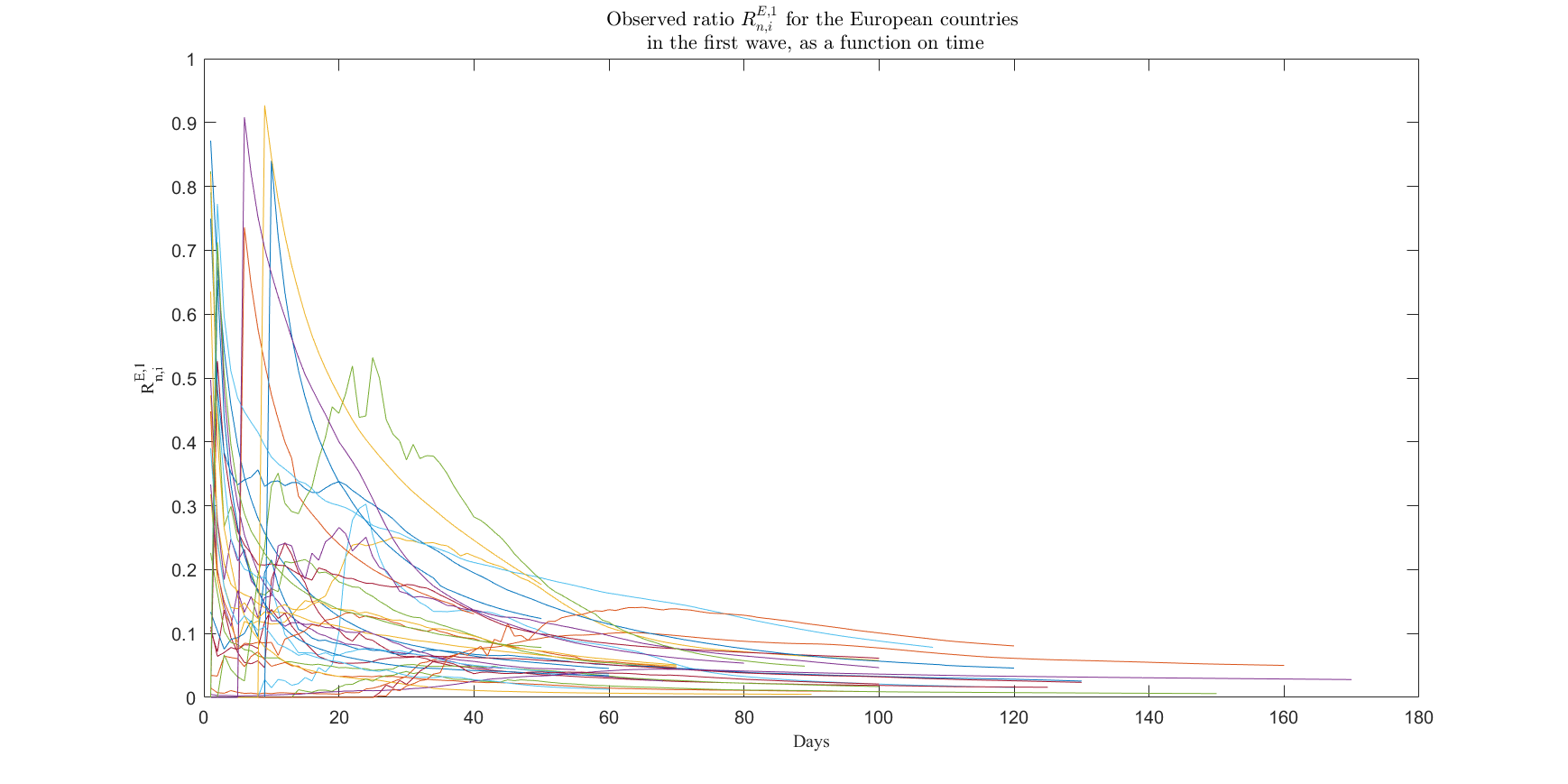}}
\caption{ }
\end{figure}

\begin{figure}[H]
\centering
\center{\includegraphics[width=\textwidth]
        {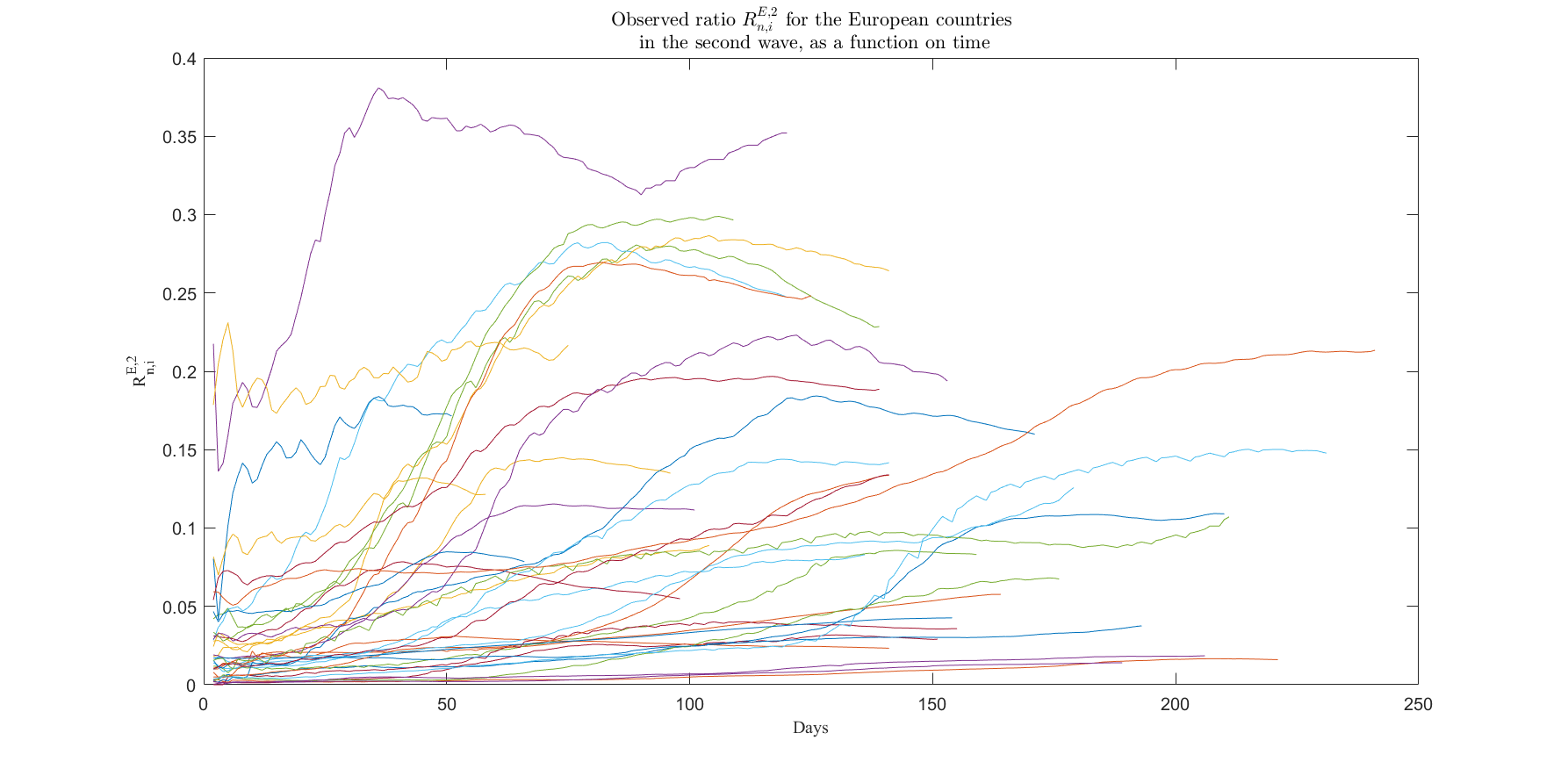}}
\caption{ }
\end{figure}

\section{Acknowledgement}
The author would like to thank Dr.~Maurizio Crippa for  helpful discussion about insight of the present paper, and R.~Gioia, C.E.O of H-DATA S.r.l.s. for technical support for the graphical elaborations.

\bibliographystyle{unsrt}




\end{document}